\documentclass[showpacs,floatfix,nofootinbib]{revtex4}

\usepackage{amsmath,amssymb}
\usepackage[dvips]{graphicx}

\newcommand{\be}{\begin{equation}}
\newcommand{\ee}{\end{equation}}
\newcommand{\nablab}{\boldsymbol{\nabla}}
\newcommand{\IF}{I}

\newcommand{\Tu}{T_{\!\boldsymbol{u}}}
\newcommand{\uu}{\boldsymbol{u}}

\newcommand{\rr}{\boldsymbol{r}}

\newcommand{\Pu}{\boldsymbol{P}_{\!\!\boldsymbol{u}}}
\newcommand{\Pw}{\boldsymbol{P}_{\!\boldsymbol{w}}}
\newcommand{\Pc}{\boldsymbol{P}_{\!\!c}}

\newcommand{\Su}{\mathcal{S}_{\boldsymbol{u}}}

\newcommand{\eq}[1]{(\ref{#1})}
\newcommand{\eqs}[1]{(\ref{#1})}

\begin{document}

\title{Variational approach to dequantization}

\author{Ricardo A. Mosna}
\email{mosna@ime.unicamp.br} \affiliation{Departamento de Matem\'atica,
Universidade Estadual de Campinas, CP 6065, 13083-859, Campinas, Brazil.}

\author{I. P. Hamilton}
\email{ihamilto@wlu.ca} \affiliation{Department of Chemistry,
Wilfrid Laurier University, Waterloo, Canada N2l 3C5.}

\author{L. Delle Site}
\email{dellsite@mpip-mainz.mpg.de} \affiliation{Max-Planck-Institute
for Polymer Research, Ackermannweg 10, D 55021, Mainz, Germany.}



\date{\today}

\begin{abstract}
We present a dequantization procedure based on a variational
approach whereby quantum fluctuations latent in the quantum momentum
are suppressed. This is done by adding generic local deformations to
the quantum momentum operator which give rise to a deformed kinetic
term quantifying the amount of ``fuzzyness'' caused by such
fluctuations. Considered as a functional of such deformations, the
deformed kinetic term is shown to possess a unique minimum which is
seen to be the classical kinetic energy. Furthermore, we show that
extremization of the associated deformed action functional
introduces an essential nonlinearity to the resulting field
equations which are seen to be the classical Hamilton-Jacobi and
continuity equations. Thus, a variational procedure determines the
particular deformation that has the effect of suppressing the
quantum fluctuations, resulting in dequantization of the system.

\end{abstract}

\maketitle


Quantum mechanics is an extremely successful theory for the description
of atomic and molecular systems.
Its predictions of microscopic phenomena are highly accurate and it is unrivalled
as a physical theory. On the other hand, because of the undisputable success of
classical mechanics in its domain of validity, there is continued
interest in dequantization procedures whereby the classical regime
is obtained from the quantum one. Here by dequantization we do not
mean the procedure of obtaining a semiclassical limit of a given
quantum system, as in the WKB approximation. Rather, following
\cite{AGM}, by dequantization we mean ``a set of rules which turn
quantum mechanics into classical mechanics''.

An insightful step towards dequantization is the introduction of
formulations of classical mechanics that are operator based. The
earliest such formulation is that of Koopman \cite{KM} and von
Neumann \cite{NM}. These works were the foundation of more recent
path integral formulations of classical mechanics \cite{GRT} and the
related dequantization procedure of Abrikosov, Gozzi and Mauro
\cite{AGM}. On the other hand, there has been interest in
quantization procedures formulated in a quasi-classical language,
whereby stochastic terms are added to the equations of classical
mechanics. In particular, Nelson \cite{nelson} and earlier work of
F\'enyes \cite{FE} and Weizel \cite{WE} showed that the
Schr\"odinger equation can be derived from Newtonian mechanics via
the assumption that a classical particle is subjected to Brownian
motion with a real diffusion coefficient. Also, Hall and Reginatto
\cite{HR} have shown that the Schr\"odinger equation can be derived
from the classical equations of motion by adding fluctuations obeying an exact
Heisenberg-type equality to the classical momentum. In a similar
vein, Reginnato \cite{reginnato} has shown that the Schr\"odinger
equation can be derived by minimization of the Fisher information~\cite{fisher}.

In this Letter we present a dequantization procedure whereby
classical mechanics is derived from quantum mechanics by suppressing the
effects of such ``quantum fluctuations''. To develop this approach
within a consistent mathematical framework, we introduce local
deformations of the momentum operator, which correspond to
fluctuations of the quantum momentum. These naturally induce an
associated deformed kinetic term, which quantifies the amount of
``fuzzyness'' caused by these fluctuations. Considered as a
functional of such deformations, the deformed kinetic term is shown
to possess a unique minimum which is seen to be the classical
kinetic energy. Furthermore, we show that extremization of the
associated deformed action functional introduces an essential
nonlinearity to the resulting field equations which are seen to be
the classical Hamilton-Jacobi and continuity equations. The
minimizing deformation can thus be interpreted as the particular
deformation that removes the quantum fluctuations so that the
classical case, i.e., dequantization, is attained. Moreover, the
minimizing deformation automatically determines an expression for
the quantum fluctuations which, when added to the classical
momentum, leads to the quantum one. This expression is shown to be
identical to Nelson's osmotic momentum.

\section{Minimizing the deformed kinetic term}
\label{sec:minimizing}

We begin by considering a local deformation $\boldsymbol{P}\to\Pw$
of the quantum momentum operator $\boldsymbol{P}=-i\hbar\nablab$ for
an one-particle system (the generalization to many-particle systems
of scalar particles is straightforward), with
\begin{equation}
\Pw\psi = \left( \boldsymbol{P} - \boldsymbol{w} \right) \psi,
\label{Pw}
\end{equation}
where $\psi$ is the wavefunction of the system and $\boldsymbol{w}$ is a
position-dependent (complex) vector field.%
\footnote{The most general linear operator acting on $\psi$ can be written, in the position representation,
as $\psi(\rr\!)\to\psi'(\rr\!)=\int K(\rr,\rr'\!)\psi(\rr'\!)d\rr'$. The deformation considered in \eq{Pw}
is local in the sense that the value of $(\Pw-\boldsymbol{P})\psi$ at a given point $\rr$ is a function
of $\psi(\rr\!)$ only (in fact, it is given by $-\boldsymbol{w}(\rr\!)\psi(\rr\!)$).}
Since our aim is to \emph{dequantize} the system (thereby
leaving the realm of quantum mechanics), there is no {\em a priori}
reason to assume that $\Pw$ is Hermitian when $\boldsymbol{w} \ne
0$. Writing $\boldsymbol{w}=\boldsymbol{v}+i\uu$, where
$\boldsymbol{v}$ and $\boldsymbol{u}$ are respectively the real
and imaginary parts of $\boldsymbol{w}$, we see that the term
$\boldsymbol{v}$ in $\Pw\psi=-(i\hbar\nablab+ \boldsymbol{v})\psi
-i\uu\psi$ acts in the same way as an electromagnetic field
$\boldsymbol{A}$, which is known to change the quantum momentum
operator $-i\hbar\nablab$ to $-i\hbar\nablab + \kappa\boldsymbol{A}$,
where $\kappa$ is a constant.
Therefore, in what follows we restrict the deformations in \eq{Pw}
to those corresponding to imaginary $\boldsymbol{w}$, so that
$\boldsymbol{w}=i\uu$, with $\uu$ real:
\begin{equation}
\Pu\psi = \left( \boldsymbol{P} - i\uu \right) \psi.
\label{Pu}
\end{equation}

Let
\begin{equation}
T=\frac{1}{2m}\int (\boldsymbol{P}\psi)^\ast (\boldsymbol{P}\psi) d\tau
\label{Tqm}
\end{equation}
and
\begin{equation}
\Tu =\frac{1}{2m}\int (\Pu\psi)^\ast
(\Pu\psi) d\tau \label{Th0}
\end{equation}
be the kinetic terms arising from $\boldsymbol{P}$ and $\Pu$, respectively,
where $m$ is the mass of the particle and $d\tau$ denotes the associated volume element.
Integration by parts shows that one can alternatively write
\begin{equation}
T=\frac{1}{2m}\int \psi^\ast\boldsymbol{P}^2\psi d\tau, \tag{\ref{Tqm}a}
\label{Tqm_prime}
\end{equation}
as usual, and
\begin{equation}
\Tu =\frac{1}{2m}\int \psi^\ast \Pu^\dag\Pu\psi d\tau, \tag{\ref{Th0}a}
\label{Th0a}
\end{equation}
where
\[
\Pu^\dag\psi = \left( \boldsymbol{P} + i\uu \right) \psi
\]
is the adjoint of $\Pu$. Note that although $\Pu$ and $\Pu^\dag$
are, in general, not Hermitian operators, $\Pu^\dag\Pu$ is always
Hermitian so that $\Tu$ (like $T$) is always a real quantity.

We then have
\begin{equation}
\Tu = T + \frac{1}{2m} \int \rho \left( -\hbar\nablab\cdot\uu + \lVert\uu\rVert^2 \right) d\tau,
\label{Tu}
\end{equation}
where $\rho=\psi^*\psi$.
Note that $\Tu=\Tu[\psi,\boldsymbol{u}]$ is a
functional of both $\psi$ and $\boldsymbol{u}$. Therefore, the
full-fledged variational principle associated with $\Tu$ should
involve minimization with respect to both $\psi$- and
$\boldsymbol{u}$-variations.

A straightforward calculation shows that variation of
$\Tu[\psi,\boldsymbol{u}]$ with respect to $\boldsymbol{u}$ yields
\begin{equation}
\frac{\delta \Tu}{\delta \uu} = \frac{1}{2m}(2\rho\uu + \hbar\nablab\rho).
\label{delT}
\end{equation}
Therefore, extremization of
$\Tu$ with respect to $\boldsymbol{u}$-variations leads to the
critical point
\begin{equation}
\uu_c = -\frac{\hbar}{2}\frac{\nablab\rho}{\rho}.
\label{umin}
\end{equation}
This in turn corresponds to the deformed momentum operator
\begin{equation}
\Pc \psi = \left( \boldsymbol{P} + \frac{i\hbar}{2}\frac{\nablab\!\rho}{\rho} \right) \psi,
\label{Pc}
\end{equation}
which has been introduced in \cite{clasquan} from a different
perspective (we come back to this point later).

Expanding $\Tu$ around the critical point yields
\be
T_{\uu_c+\delta\boldsymbol{u}} = T_{\uu_c} + \frac{1}{2m} \int \rho \,
\lVert\delta\uu\rVert^2 d\tau,
\ee
which shows that the deformation $\uu_c$ of \eq{umin} leads to the {\em unique minimum}
of $\Tu$, given by
\be
T_{\uu_c}=T-\frac{\hbar^2}{8m}\IF,
\label{Tumin}
\ee
where $\IF$ is the Fisher information \cite{fisher},
\begin{equation}
\IF= \int \frac{\left( \nablab\!\rho
\right)^2}{\rho} d\tau.
\label{fisher}
\end{equation}

\begin{figure}[htbp]
\begin{center}
\includegraphics[width=7cm]{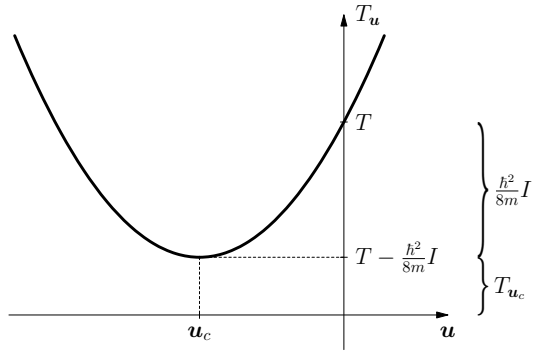}
\caption{The minimum of $\Tu[\psi,\uu]$ under $\uu$-variations is attained for
         $\uu_c=-\frac{\hbar}{2}\frac{\nablab\rho}{\rho}$, with $T_{\uu_{\!c}}=T-\frac{\hbar^2}{8m}\IF$.
         The plot illustrates this point by pictorially representing the
         (infinite-dimensional) $\uu$-space on its abscissa.}
\label{fig:minimum}
\end{center}
\end{figure}

Thus we have shown that there is a unique solution to the
deformation parameter $\uu$ which minimizes the deformed kinetic
term $\Tu$ under $\uu$-variations. For a pictorial representation
of the $\uu$-dependence of $\Tu$, see Fig.~\ref{fig:minimum}.

A straightforward calculation shows that the action of $\Pc$ on
the wavefunction $\psi=\sqrt{\rho}e^{iS/\hbar}$ of the system is
given by
\begin{equation}
\Pc\psi=\nablab S \: \psi, \label{Pcl1}
\end{equation}
so that, from \eq{Th0},
\begin{equation}
T_{\uu_c}=\frac{1}{2m} \int \rho \, \lVert\nablab S\rVert^2 d\tau.
\label{Tuc2}
\end{equation}
This is exactly the mean kinetic energy of a classical ensemble,
described by the density $\rho$, with associated Hamilton's
principal function $S$ and momentum field $\nablab S$
\cite{Goldstein,Holland}. We therefore refer to $T_{\uu_c}$ as the
classical kinetic term associated with the ensemble defined by $\rho$
and $S$.%
\footnote{We note that \eq{umin} can be rewritten as $\uu_c=\hbar \nablab\! f$,
with $f=-\ln(\rho)/2$, which is precisely the expression for the
deformation function that was introduced in~\cite{clasquan} in an
\emph{ad hoc} way. It is now clear that this expression has a deeper
justification, since it corresponds to the \emph{unique} momentum
deformation which \emph{minimizes} the deformed kinetic term.}

\section{From quantum to classical}
\label{sec:dequant}

Recall that the Schr\"odinger equation can be derived from the action functional
\begin{equation}
\mathcal{S}= \int\left[ i\frac{\hbar}{2}\left(
\psi^{\ast}\frac{\partial\psi}{\partial t}-
\frac{\partial\psi^{\ast}}{\partial t}\psi\right) -
\frac{1}{2m}(\boldsymbol{P}\psi)^{\ast}\cdot(\boldsymbol{P}\psi)-V\psi^{\ast}\psi\right]
d\tau dt
\label{Ssch}
\end{equation}
through the usual variational procedure associated with the fields $\psi$ and $\psi^\ast$.
The deformation of the momentum operator $\boldsymbol{P}\to\Pu$ has the effect of
introducing a deformation $\mathcal{S}\to\Su$, resulting in
\begin{align}
\Su &= \int\left[ i\frac{\hbar}{2}\left( \psi^{\ast}\frac{\partial\psi}{\partial t}-
       \frac{\partial\psi^{\ast}}{\partial t}\psi\right) -
       \frac{1}{2m}(\Pu\psi)^{\ast}\cdot(\Pu\psi)-
       V\psi^{\ast}\psi\right] d\tau dt     \notag          \\
    &= -\int \Tu[\psi,\uu] \, dt+
       \int\left[ i\frac{\hbar}{2}\left( \psi^{\ast}\frac{\partial\psi}{\partial t}-
       \frac{\partial\psi^{\ast}}{\partial t}\psi\right) -
       V\psi^{\ast}\psi\right] d\tau dt,     \label{Sh2}
\end{align}
which depends on $\uu$ only through its first term.

This yields a deformed action $\Su[\psi,\uu]$ whose
field equations are obtained by varying both $\psi$ and $\uu$.
From our previous discussion, we see by the form of
\eq{Sh2} that extremization with respect to $\uu$-variations fixes
$\uu$ to be given by \eq{umin}. Substituting in \eq{Sh2} results in a
reduced action, depending only on $\psi$, of the form
\be
\mathcal{S}_c[\psi]= \int\left[ i\frac{\hbar}{2}\left(
\psi^{\ast}\frac{\partial\psi}{\partial t}-
\frac{\partial\psi^{\ast}}{\partial t}\psi\right) -
\frac{1}{2m}(\Pc\psi)^{\ast}\cdot(\Pc\psi)-
V\psi^{\ast}\psi\right] d\tau dt, \label{Scl}
\ee
with $\Pc$ given by \eq{Pc}. As discussed in~\cite{clasquan}, \eq{Scl} is precisely
the classical version of the action~(\ref{Ssch}). This can be
most easily seen by noting that, when $\psi$ is written in polar
form, $\psi=\sqrt{\rho}e^{iS/\hbar}$, \eq{Scl} becomes
\begin{equation}
\mathcal{S}_c[\rho,S]=-\int\rho\left( \frac{\partial S}{\partial t}+
                      \frac{\left( \nablab S\right) ^{2}}{2m}+
                      V\right) d\tau dt,
\label{Scl2}
\end{equation}
which gives rise, upon variation of $\rho$ and $S$, to the
classical Hamilton-Jacobi and continuity equations%
\footnote{Alternatively, we could have directly calculated the Euler-Lagrange equations associated
with the deformed action $\Su[\psi,\uu]$. The Euler-Lagrange equations relative to $\uu$
then yield $2\rho\uu+\hbar\nablab\rho=0$, while the Euler-Lagrange equations relative to
$\psi$ and $\psi^\ast$ yield
$i\hbar\frac{\partial\psi}{\partial t}=-\frac{\hbar^2}{2m}\nabla^2\psi+
\left[ V+\frac{1}{2m}\left( \uu^2\! - \hbar\nablab\!\!\cdot\!\uu \right) \right]\psi=0$.
Isolating $\uu$ from the former equation and substituting into the latter
then yields, in terms of $\rho$ and $S$, \eqs{eqHJ} and (\ref{eqcont}).}
\begin{align}
\frac{\partial S}{\partial t} + \frac{(\nablab\! S)^2}{2m}+V & = 0, \label{eqHJ} \\
\frac{\partial\rho}{\partial t} + \nablab\cdot\left( \rho \frac{\nablab\! S}{m}\right) & = 0.  \label{eqcont}
\end{align}

Thus we have shown that the procedure of ``deforming the momentum
and extremizing its associated deformed action functional'' can be
effectively regarded as a {\em dequantization method}, at least for
scalar particles. Note that the choice of deformed momentum that
extremizes its associated deformed action is given by $\Pc$ of
\eq{Pc}. This can be thought of as a classical version of the
momentum operator since its associated kinetic term, $T_{\uu_c}$, is
the mean kinetic energy of the classical ensemble defined by $\rho$
and $S$ (as discussed earlier), and its associated
deformed action, $\mathcal{S}_c$, yields the classical
Hamilton-Jacobi and continuity equations. Further justification on
the interpretation of $\Pc$ as a classical version of the momentum
operator can be found in \cite{clasquan}, where $\Pc$ was introduced
from a different perspective and shown to be equivalent to an expression
introduced by Hall in \cite{hall}
that gives the best classical estimate of the momentum which is
compatible with simultaneous knowledge of the position of the
system.

In the quantization procedure of Nelson \cite{nelson} a classical
particle is subjected to Brownian motion. In addition to its
classical velocity, a Brownian particle has a velocity due to the
osmotic force, that Nelson terms the osmotic velocity (which is half
the difference between the forward and backward drift velocities).
From Einstein's theory, the osmotic velocity is given by {\bf $\nu
\nablab \rho/\rho$} where $\nu$ is the diffusion coefficient. Since
macroscopic bodies do not appear to be subjected to Brownian motion,
Nelson assumes that $\nu$ is inversely proportional to the particle
mass and makes the ansatz $\nu=\hbar/2m$. Then the corresponding
osmotic momentum, which is the term added to the classical momentum
to give the quantum one, is ($\hbar$/2){\bf $\nablab \rho/\rho$}.
This expression is seen to be identical to minus our $\uu_c$ of \eq{umin}. This
is no coincidence and can be qualitatively understood as follows. In
Nelson's quantization approach, quantum fluctuations (expressed as
the osmotic momentum) are explicitly added to $\Pc$, thereby
resulting in the quantum momentum. In our dequantization approach,
the quantum fluctuations latent in $\boldsymbol{P}$ are stripped off
by the process of minimizing $\Tu$, thereby isolating the classical
momentum. In the process, our dequantization approach
\emph{automatically} identifies the expression for $-\uu_c$ (cf
\eq{umin}).

\section{Discussion}

We have presented a dequantization procedure based on a
variational principle whereby quantum fluctuations latent in the
quantum momentum are suppressed. To this end, we added generic local
deformations $\uu$ to the quantum momentum operator
$\boldsymbol{P}$. Such deformations are independent of $\psi$ and
consequently the deformed momentum operator $\Pu$ is linear in
$\psi$ (cf \eq{Pu}). However, \emph{after} extremization of the
associated deformed kinetic term $\Tu$, $\uu$ becomes {\em
dependent} on $\psi$ (and fixed to
$\uu=-\frac{\hbar}{2}\frac{\nablab\rho}{\rho}$), giving rise to the
{\em nonlinear} classical momentum operator $\Pc$ of \eq{Pc}.
Furthermore, extremization of the associated deformed action, $\Su$,
gives rise to the classical Hamilton-Jacobi and continuity equations
(\eqs{eqHJ} and (\ref{eqcont})), so that dequantization is attained.

For a classical system described by a probability density there is
uncertainty in the position (and momentum). For the corresponding
quantum system there is additional uncertainty. The dequantization
method presented here removes the additional part of the uncertainty
that is quantum leaving only the uncertainty that is classical and it
does this in a ``minimalist" way --- without introducing any artifacts
--- through a deformation procedure based on a variational principle.
As a result of the dequantization procedure the quantum momentum
fluctuations are suppressed and, in this sense, the momentum-space
localization of the system (thereupon considered as a classical
ensemble) is increased. However, the spatial
localization of the system is unchanged as this quantity is
determined by $\rho$ which is unaffected by the dequantization
process (in fact, as noted above, $\rho$ determines the deformation
function). For a system which is more spatially localized both the
quantum kinetic term and the Fisher information are larger and in
the limit of extreme spatial localization both become infinite but
the classical kinetic term can remain finite.

The approach presented here may shed light on the quantum-classical
transition, since the passage from the {\em linear} equations of
quantum mechanics to the {\em nonlinear} equations of classical
mechanics is made salient through the deformation function
$\boldsymbol{u}$. A remarkable effect of this linearity {\em vs}
nonlinearity issue is the fundamentally different characterization
of the concept of chaos in quantum and classical systems
\cite{gutzwiller}. Consideration of the deformed action $\Su$ for
generic fixed values of $\boldsymbol{u}$, different from
$\boldsymbol{0}$ and $\boldsymbol{u}_c$,  may contribute novel
insights in this regard. It may also facilitate novel formulations
of semiclassical mechanics.

As noted previously, dequantization is ``a set of rules which turn
quantum mechanics into classical mechanics" \cite{AGM}. For the
dequantization procedure proposed here these rules are ``deform the
momentum and extremize its associated deformed action functional''.
It would be interesting to investigate how this procedure (developed
here for scalar particles) can be extended to other contexts, such
as particles with internal degrees of freedom.

\acknowledgments

We thank  B. D\"unweg, E. Gu\'eron, M. J. W. Hall and M. Reginatto for helpful comments.
RAM acknowledges FAPESP for financial support. IPH acknowledges
funding from NSERC and thanks Wilfrid Laurier University and the
Fields Institute for support.


\begin{thebibliography}{99}

\bibitem{AGM} A. A. Abrikosov Jr, E. Gozzi and D. Mauro, Ann. Phys.  {\bf 317}, 24 (2005).

\bibitem{KM} B. O. Koopman, Proc. Nat. Acad. Sci. USA {\bf 17}, 315 (1931).

\bibitem{NM} J. von Neumann, Ann. Math. {\bf 33}, 587 (1932).

\bibitem{GRT} E. Gozzi, M. Reuter and W.D. Thacker, Phys. Rev. D {\bf 40}, 3363 (1989);
 {\bf 46}, 757 (1992).

\bibitem{nelson} E. Nelson, Phys. Rev. {\bf 150}, 1079 (1966);
E. Nelson, \textit{Dynamical Theories of Brownian Motion} (Princeton Univ. Press, Princeton, 1967).

\bibitem{FE} I. F$\acute{\rm e}$nyes, Z. Physik {\bf 132}, 81 (1952).

\bibitem{WE} W. Weizel, Z. Physik {\bf 134}, 264 (1953); {\bf 135}, 270 {1953}; {\bf 136}, 582 (1954).

\bibitem{HR} M. J. W. Hall and M. Reginatto, J. Phys. A {\bf 35}, 3289 (2002), quant-ph/0102069.

\bibitem{reginnato} M. Reginatto, Phys. Rev. A {\bf 58}, 1775 (1998).

\bibitem{fisher} R. A. Fisher, Proc. Cambridge Philos. Soc. {\bf 22}, 700 (1925).

\bibitem{clasquan} R. A. Mosna, I. P. Hamilton and L. Delle Site, J. Phys. A {\bf 38}, 3869 (2005), quant-ph/0504124.

\bibitem{Goldstein} H. Goldstein, \textit{Classical Mechanics, 2nd ed.} (Addison-Wesley, Reading, MA, 1980).

\bibitem{Holland} P. R. Holland, \textit{The Quantum Theory of Motion} (Cambridge University Press, Cambridge, 1993).

\bibitem{hall} M. J. W. Hall, Phys. Rev. A {\bf 62}, 12107 (2000),
quant-ph/9912055; M. J. W. Hall, Phys. Rev. A {\bf 64}, 52103
(2001), quant-ph/0107149.

\bibitem{gutzwiller} See, {\em e.g.}, M. C. Gutzwiller, \textit{Chaos in Classical and Quantum Mechanics}
(Springer-Verlag, New York, 1990).


\end{thebibliography}
\end{document}